\documentclass{llncs}
\usepackage{xcolor}
\usepackage{soul}

\title{Of Hags and bitches. Ageist attitudes in 2016 presidential debate on twitter}
\author{Bartłomiej Balcerzak, Radosław Nielek}
\institute{Polish-Japanese Academy of Information Technology, Warsaw 02008, Poland,\\
\email{b.balcerzak@pjwstk.edu.pl, nielek@pjwstk.edu.pl}}
\begin{document}
\maketitle

\begin{abstract}
In this article we present our exploratory research into the occurrence of ageist attitudes within the discussion related to the US 2016 presidential election. We use natural processing techniques to analyze the content tweets related to Hillary Clinton and Donald Trump. Content analysis shows that although ageist attitudes are scarce in the discussion, they are mostly focused on Hillary Clinton rather than Donald Trump. Also, ageist arguments against Donald Trump appear mostly as a reply to controversies connected with the health of Hillary Clinton.
\end{abstract}

\section{Introduction}

In this article we present how ageist attitudes can manifest in social media. As an example we used tweets connected with the US 2016 presidential campaign. The choice for this particular set of data was based on two main reasons. Firstly, both presidential candidates are considerably older than the current president of the US. Both are older than than 65 years old, with Hillary Clinton being 69 and Donald Trump being 70, which also makes them the oldest presidential candidates in US history. Secondly due to the controversies raised during the current presidential campaign it is possible to compare how ageist attitudes compare to other topics that are raised in relation to the presidential candidates.
Ageism is a form of social discrimination in which people mostly older adults are subject to prejudice and unfair treatment based on their age alone\cite{butler1980ageism}. Just like other forms of social exclusion (such as racism and sexism), ageism is an obstacle in creating a fair and inclusive society.
Our research is mostly exploratory in nature, we want to learn what kind of ageist attitudes are apparent in relation to Hillary Clinton and Donald Trump. We also proposed two research objectives:
\begin{itemize}
\item See if the presidential candidates are subject to ageist attitudes and comments, and if so, which of the candidates is more often the target of such comments.
\item Research if ageism in the tweets is connected with other forms of discriminatory or offensive language (ie. racist or sexist).
\end{itemize}
By setting these objectives we want to research the interaction between objective age, and the perception of somebody's age, as well as the relation ageism has to other forms of social exclusion.
\section{Related work}
In order to research the occurrence of ageist attitudes in social media, one needs to setup up a framework for how ageism is represented in language. Such a comprehensive framework of ageism has been proposed by \cite{gendron2015language}. In their work they present a multidimensional model for representing ageist attitudes in language. These dimension varied from negative sentiment towards older adults, through judgments and stereotypical assumptions, to infantilization of older adults when creating a narrative. Within this paper we focus mostly on one of the aspects of this framework, which is negative associations with being old.
In studies related to the perception of older adults it has been stated that such negative associations play an important role. Work by \cite{connolly2016college} where perception of aging among college students were tested are a prime example of that.
Other works related to representation of older adults in media, and social media were focused on content analysis (such as work done by \cite{brooks2016s} where a content analysis of the US Superbowl commercials was made in order to review how they represent older adults), or on the potential for social advocacy \cite{trentham2015social} 
Automated detection of social bias in social media has been attempted mostly in the field of detecting hate speech \cite{warner2012detecting} and racism against minorities \cite{kwok2013locate}. Those approaches were based on machine learning and natural language processing, with the aim of effective detecting of racism in social media. Most of the research was done on Twitter. To the best of the authors' knowledge no similar approach to detecting ageism has been proposed so far.
\section{Methodology}
\subsection{Data set}
We collected over 300 000 tweets in English, which contained any reference to either of the presidential candidates in the 2016 campaign: Donald Trump, and Hillary Clinton. In order to collect those tweets we prepared a code with the use of Python Tweepy library, which allows for automated use of the Twitter streaming API. The tweets were collected during the last week of August.
Table \ref{corpus} presents a more detailed description of the content of the corpus. The numbers of tweets relating to Hillary Clinton and Donald Trump are comparable in number, however the Trump tweets are more numerous.
\begin{table}[!h]
\centering
\caption{Corpus parameters}
\label{corpus}
\begin{tabular}{|c|| c| c| c|}
\hline
 &Total tweets & Clinton tweets & Trump tweets\\
 \hline
Number of tweets & 332228 & 141580 & 190648\\
Number of words & 4094 771 & 1306789 & 2787982\\
\hline
\end{tabular}
\end{table}
\newline

\subsection{Analysis tools}
In order to analyze the collected materials we wrote a set of scripts in python with the use of NLTK (version 3.2.1) library. Firstly we created a word list representing the frequency of words used in the corpus. We limited the list to words that appear more the 5 times in the corpus, and are longer than two characters. We also removed English stop words. Later we generated separate word lists for two of the candidates in order to compare which words appear more frequently in relation to each candidate. Afterwards we selected offensive words out of which we created a lexicon. We selected all of the tweets that contained at least one of those words and manually tagged them as either targeting Hillary Clinton or Donald Trump. In total 679 occurrences were observed and tagged.
\section{Results}
\subsection{Lexical representations of ageism}
In our manual review of the word list for all tweets we found only one word that clearly conveys a negative attitude towards older adults, was the word 'hag'. As shown in table, the word appeared only 36 times, which is extremely rare. When compared with other derogatory terms such as 'moron', 'dumb' or 'bitch' and 'idiot' this word also seems to be rare.
\begin{table}[!h]
\centering
\caption{Offensive words occurrence}
\label{lexical}
\begin{tabular}{|c|| c| c| c|}
\hline
Word &Total tweets & anti Clinton tweets & anti Trump tweets\\
 \hline
\textbf{Hag} & \textbf{37} & \textbf{37} & \textbf{0}\\
Moron & 64 & 22 & 42\\
Idiot & 168 & 55& 113\\
Dumb & 193 & 44& 147\\
Bitch & 219 & 108& 111\\
\hline
\end{tabular}
\end{table}
\newline
However, when the word occurrence is compared between the word lists of tweets referring to both candidates, the word 'hag' appears only in tweets that target the democratic candidate. What is more this is the only term exclusive to attacks on Hillary Clinton. It is noteworthy that a gendered term 'bitch' appears with almost equal frequency in tweets relating to both candidates.
What is also noteworthy, is the fact that although Donald Trump is the older one among the presidential candidates, tweets at attack him mostly are focused on his intelligence.
When analyzing the word list we also tested some other candidate words, that would suggest ageist attitudes. The prime candidates were words 'age', and 'old'. However, upon analysis of bigrams and trigrams containing these word, the context in which these word appeared was related mostly to matters other than age (such as old emails by Hillary Clinton, and old tweets made by Donald Trump).
\section{Conclusion}
In order to research the occurrence of ageist attitudes within the 2016 US presidential election we decided to analyze the word frequency in tweets related to both presidential candidates. While in our research we found that ageist terminology and attitude are extremely rare in the narration of the twitter users interested in the presidential campaigns a single important trend can be observed. Ageist comments and offensive remarks were aimed exclusively at the democratic candidate, Hillary Clinton. The most visible example of this trend was referring to her as a 'hag', or an 'old witch'. These results are very intriguing, when one considers the fact that Hillary Clinton is the younger of the the two candidates. The fact that Hillary Clinton is the target of ageist remarks on Twitter, as well as the fact that the word most often used in those remarks is a gendered term 'hag' while another gendered slur 'bitch' is represented equally in tweets attacking both presidential candidates, leads to a suggestion that ageist attitudes can exist independently from other forms of discriminatory language.
\section{Application of findings for social media design for older adults}
Our findings suggest that when applying design principles for social media for older adults, consideration has to be made for the influence of various forms of discriminatory language. Since, ageist attitudes can exists both independently and in conjunction with other forms of discrimination, the designer should include experiences of users representing various identities. This suggests that user experience can be a viable tools for designing social media for older adults.
\section{Further work}
In our future work we plan to extend our research into detecting ageist attitudes. We still aim at the use of data from social media, including Twitter. Our main objective is to create an unsupervised model for mapping and detecting ageist attitudes on-line. We plan to utilize the framework created by \cite{gendron2015language} in order to build a model that would apply their theoretical findings into a computer algorithm for screening ageist materials in social media.

\subsubsection*{Acknowledgments.}
This project has received funding from the European Union’s Horizon 2020 research and innovation programme under the Marie Skłodowska-Curie grant agreement No 690962.

\bibliography{biblo}
\bibliographystyle{abbrv}

\end{document}